\documentclass[%
 aip,
 amsmath,amssymb,
reprint,%
]{revtex4-1}
\usepackage{graphicx}  
\usepackage{dcolumn}   
\usepackage{bm}        
\usepackage{amssymb}   
\usepackage{amsmath}   
\usepackage{comment} 
\usepackage[caption=false]{subfig}
\usepackage{wrapfig}
\usepackage{float}
\usepackage{adjustbox}
\usepackage{multirow}
\usepackage{tabularx}
\usepackage{caption}
\usepackage{wrapfig}
\usepackage{color,soul}
\usepackage{afterpage}

\newcolumntype{Y}{>{\centering\arraybackslash}X}

\usepackage{tikz}
\begin{document}
\preprint{AIP/123-QED}
\title{
Diamagnetic to paramagnetic oscillations in an exploding Coulomb gas
}
\author{O. Zandi}\email{omid.zandique@gmail.com}
\affiliation{Department of Chemistry, University of Illinois at Urbana-Champaign}
\affiliation{Materials Research Laboratory, University of Illinois at Urbana-Champaign}
\author{R. M. van der Veen}
\affiliation{Department of Chemistry, University of Illinois at Urbana-Champaign}
\affiliation{Materials Research Laboratory, University of Illinois at Urbana-Champaign}
\affiliation{Department of Materials Science and Engineering, University of Illinois at Urbana-Champaign}
\author{P. M. Duxbury}
\affiliation{Department of Physics and Astronomy, Michigan State University}
\author{B. S. Zerbe}\email{zerbe@msu.edu}
\affiliation{Department of Physics and Astronomy, Michigan State University}
\date{\today}
\begin{abstract}
\end{abstract}
\maketitle

Dense positively-charged plasmas generated through the irradiation of particles
with high intensity ultrashort laser pulses
has been a topic of considerable experimental and theoretical research within the
past two decades\cite{Kaplan:2003_shock,Kovalev:2005_kinetic_spherically_coulomb_explosion,Last:1997_analytic_coulomb_explosion,Murphy:2014_cold_ions,Degtyareva:1998_gaussian_pileup}.
Such ultrashort laser
pulses can result in the effective removal of 
electrons from the particles, a process called laser
ablation, and if unconfined, 
the resulting positively ionized particles left behind
undergo rapid expansion.  
Such dynamics has come to be known as
Coulomb explosion within the literature\cite{Last:1997_analytic_coulomb_explosion,Grech:2011_coulomb_explosion}.
Analogous Coulomb explosion dynamics can be observed within the probe used in ultrafast
electron microscopy (UEM) where a rapidly emitted electron bunch undergoes
similar Coulomb-driven free-expansion\cite{Luiten:2004_uniform_ellipsoidal,Siwick:2002_mean_field,Reed:2006_short_pulse_theory,Zerbe:2018_coulomb_dynamics}.
The primary difference between electron and ion dynamics is in their time scale.
That is, as electrons are
far lighter than ionized atoms or molecules, the dynamics evolve
within the ultrafast regime; 
otherwise, the mathematical description is analogous.

Recently, Coulomb explosion dynamics of an electron 
single component plasma (SCP) generated
within the objective lens magnetic field of a UEM device
have been witnessed by both our group\cite{Zandi:2020_lensing} and others\cite{Sun:2020_transition}.
Due to the presence of the magnetic field, the
size of the Coulomb exploding bunch transverse to the magnetic field oscillates while it
continues to 
expand in the longitudinal direction along the magnetic field.
We presented a very simple,
non-interacting model that adequately captured the dynamics we witnessed
in our experiment provided we fit the initial 
Pearson correlation, or chirp, between the transverse
velocity and spatial components to the data;
this was justified as such a chirp arises in Coulomb explosions.
Operating at higher densities and strong magnetic confinement, 
Sun \textit{et al} witnessed evolution of the 
oscillation frequency in the transverse oscillations\cite{Sun:2020_transition};
namely,
as the expanding electron SCP began to
interact with the containment device, the plasma
shifted from lower to upper hybrid modes, which
are plasma normal modes shifted from the cyclotron
frequency due to space-charge effects\cite{Jeffries:1983_space_charge_shift,Heinzen:1991_measurement,Dubin:1993_equilibrium}.

However, the Coulomb explosion within a magnetic field differs
from the situations treated in the trapped plasma literature.
Trapped plasmas, especially SCPs, 
can be maintained for long periods of time\cite{Dubin:1996_correlations,Danielson:2015_review}. 
Such
plasmas can be described by a rigid-rotor equilibrium where
the plasma particles within the plasma collectively rotate 
similar to a rigid-body\cite{Bogema:1970_rigid_rotor} with an equilibrium distribution
determined by the angular momentum and temperature of the plasma\cite{Davidson:1969_equilibrium}. 
The trapped plasma relatively rapidly approaches such rigid-rotor equilibrium,
and therefore substantial attention has focussed on the description of plasma 
dynamics near this thermal equilibrium\cite{Jeffries:1983_space_charge_shift,Dubin:1991_letter,ONeil:1998_thermal}.
The period of time during which the plasma is far-from-equilibrium is not 
as well understood, and the partially unconfined Coulomb explosion is much
better described as far-from-equilibrium than is the standard in
trapped plasmas.

Nonetheless, there is some understanding of this rapid period of thermalization.
Trapped plasmas are known to cool as they approach rigid-rotor equilibrium
due to cyclotron radiation\cite{ONeil:1980_cooling,Amoretti:2003_measurement}.
The RMS emittance of SCPs used for accelerator beams is known to 
increase and nearly saturate within $1/4$ the plasma period of the initial 
distribution\cite{Struckmeier:1984_stability}, and this change can be attributed to 
free-energy released from the relaxation of a beam from a non-uniform profile
to a nearly-uniform profile\cite{Wangler:1985_emittance_relaxation,Reiser:1991_free_energy}.
Disorder induced heating (DIH) exhibits similar dynamics to RMS emittance evolution\cite{Maxson:2013_DIH}
although the origin of the free-energy in DIH has been described as
quantum or stochastic\cite{Gericke:2003_dih,Murillo:2006_dih}.
Models including additional stochastic effects\cite{Struckmeier:1994_fokker_planck,Struckmeier:1996_entropy,Struckmeier:2000_stochastic,Gao:2000_langevin},
chaotic effects\cite{Kandrup:2004_frozen}, and non-linear field effects\cite{Jensen:2014_surface_emittance,Dowell:2009_analytic_emittance,Gevorkyan:2018_roughness} have described addittional
pieces of similar thermalization processes,
and these descriptions inform the use of RMS envelope equations that
are widely used by accelerator physicists to model the dynamics
of nearly uniform SCPs(for instance\cite{Lund:2014_emittance_def}).
However, as it is known that the equilibrium distribution significantly
deviates from uniform, it is not apparent whithin the literature to what 
extent a uniform distribution can capture the dyanmics of Coulomb-explosion
within a magentic field.

\begin{figure}[b!]
  \includegraphics[width=0.45\textwidth]{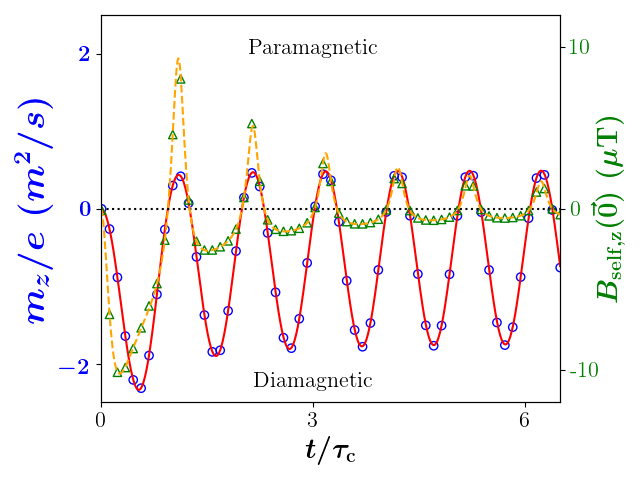}
  \caption{\label{fig:sphere:B} \small
  Diamagnetic/paramagnetic oscillation seen during Coulomb explosion of 
  an electron SCP of $10,000$ electrons distributed uniformly with an initial radius of 
  $2~\mu$m ($R^2 = 0.8~ (\mu\text{m})^2$)
  from rest in an external magnetic field of $4 T$.
  Plot shows the evolution of magnetic moment (solid red line = envelope equations; blue circles = simulation) 
  and self-magnetic field (dashed orange line = envelope equations; green triangles = simulation).
  Dotted black line at $0$ indicates when the self-magnetic effects reverse direction.
  }
\end{figure}

In our previous work, 
we developed a non-interacting, RMS envelope model.
We 
initialized this model with a non-zero Pearson correlation, or chirp, 
that we argued arose due to the action of Coulomb explosion\cite{Zandi:2020_lensing},
and we found that this model captured the dynamics adequately
for our experimental purposes.
On the other hand, the frequency shifts seen in Sun \textit{et al}'s 
experiments suggests that particle-particle interactions 
need to be considered, and when we conducted 
$N$-particle simulations of denser plasmas, we
were surprised to 
see a fundamental effect
that appears to be missing in the trapped plasma and
accelerator literatures --- transversely confined 
Coulomb explosion can lead to the self-magnetic moment of the 
electron SCP flipping its sign and driving the SCP into a paramagnetic state.
We demonstrate this effect in Fig. \ref{fig:sphere:B} where we 
show the magnetic moment
changes sign twice every period.
This signals a periodic transition between paramagnetic and diamagnetic states
for the Coulomb exploding electron SCP despite this ensemble starting with no appreciable angular momentum.

In this paper, we demonstrate this effect with $N$-particle results,
show how increasing the magnetic field effectively turns this effect on,
show how ``heating" of the electron SCP results in the elimination of the
paramagnetic portion of the oscillation, and present
three very simple models streamlined from existing accelerator literature that can
be used to understand the diamagnetic to paramagnetic oscillation
and this temperature effect.
Further, we introduce two novel physical measures: (1) a rigid-rotor statistic
and (2) an angular emittance measure.  
As we could not find a quantitative definition of rigid-rotor motion,
we present such a measure with the rigid-rotor statistic.
In the development of this statistic, we found a statistical expression that 
is nearly conserved in time for non-interacting systems and appears to be the angular complement to the radial
emittance proposed by others\cite{Lund:2014_emittance_def}.
We summarize and place these results in context of the literature in the Conclusions.

\section{$N$-particle simulations and spheroidal envelope equations from rest}

We examined the dynamics of electron SCPs exploding from rest within a uniform magnetic field
strength of $B_{ext}$
over a wide variety of initial conditions (not shown) using $N$-particle simulations,
and compared the $N$-particle results to envelope equations.
We term the direction along the magnetic field axis the longitudinal direction and the two directions
perpendicular to this direction the transverse direction(s).
$N$-particle simulations were carried out using the Velocity-Verlet algorithm\cite{Verlet:1967_algorithm} with the electric
field calculated using the Fast Multipole Method\cite{Gimbutas:2015_fmmlib3d} at every step.

Before discussing specific results, we briefly introduce our spheroidal envelope model and notation. 
For simplicity, we assumed cylindrical symmetry (perfect Pearson correlation between $x$ and $y$), with
$R^2 = \frac{1}{2} (s_x^2+s_y^2)$ representing the transverse ``radial" variance,
with $s_d^2$ representing the variance along $d$ and
$Z = s_z$ the longitudinal standard deviation; the parameters
$R$ and $Z$ are often termed the RMS envelope of the SCP as we will do here.
Under the assumption of perfect Pearson correlation between $x$ and $y$ and no Pearson correlation with $z$,
we define the radial ($\eta_R = \frac{s_{x,v_x} + s_{y,v_y}}{2 R^2}$), 
angular ($\omega_{\rm{r}} = \frac{s_{x,v_y} - s_{y,v_x}}{2R^2}$), 
and longitudinal ($\eta_Z = \frac{s_{z,v_z}}{s_z^2}$) chirps
by the line of best fit for the velocity, i.e. 
${\vec{{v}}}_{bf}(r,z) = {{\eta}}_R r \hat{R} + {{\omega}}_{\rm{r}} r \hat{\phi} + {{\eta}}_{Z} z \hat{Z}$,
where $s_{x,v_x}$ represents the covariance of $x$ and $v_x$, etc.
We use tilde above a parameter to indicate the dimensionless version of the parameter and $0$
in the subscript to indicate its initial value.
The analysis can be made dimensionless by scaling the spatial length by
the appropriate initial length and the time by the electron's cyclotron frequency
associated with the external magnetic field 
$\omega_{\rm{c}} = \frac{e B_{ext}}{m}$
where 
$e$ is the elementary charge and $m$ is the mass of the electron.
Three examples of this are
$\tilde{R} = \frac{R}{R_0}$, 
$\tau = \omega_{\rm{c}}t$,
and
${\tilde{\eta}}_R = \frac{\eta_R}{\omega_{\rm{c}}}$.
The resulting dimensionless system of equations describing the evolution of the phase
space coordinates starting from rest follows:
\begin{subequations}\label{eq:3D ddot}
\begin{align}
  \frac{d\tilde{R}}{d\tau} &= {\tilde{\eta}}_R\tilde{R}\\
  \frac{d\tilde{Z}}{d\tau} &= {\tilde{\eta}}_Z\tilde{Z}\\
  \frac{d{\tilde{\eta}}_R}{d\tau} &= \frac{1}{2} {\tilde{\omega}}_\text{\rm{p}3,0}^2 \Gamma_1\left(\alpha\right) \frac{1}{\tilde{R}^2\tilde{Z}} - \frac{1}{4} + \frac{1}{4\tilde{R}^4} - {\tilde{\eta}}_R^2\label{eq:3D omega_R}\\
 {\tilde{\omega}}_{\text{r}} &= \frac{1}{2} - \frac{1}{2{\tilde{R}}^2}\label{eq:wphi osc}\\
  \frac{d{\tilde{\eta}}_Z}{d\tau} &= \tilde{{\omega}}_\text{\rm{p}3,0}^2 \Gamma_2\left(\alpha\right) \frac{1}{\tilde{R}^2\tilde{Z}}- {\tilde{\eta}}_Z^2.\label{eq:3D Cz}
\end{align}
\end{subequations}
where 
$\omega_\text{\rm{p}3,0}^2 = \frac{e^2}{m\epsilon_0} \frac{N}{\frac{4}{3}\pi R_0^2 Z_0}$ 
is the initial plasma frequency of the SCP (in three dimensions, hence the p$3$ in the subscript),
$\alpha = \frac{Z}{R}$ is the aspect ratio, and
$\Gamma_1(\alpha) = \frac{\alpha}{\alpha^2 -1}\left(\alpha - \frac{\cosh^{-1}\alpha}{\sqrt{\alpha^2-1}}\right)$ and
$\Gamma_2(\alpha) = \frac{\alpha}{\alpha^2-1}\left(\frac{\cosh^{-1}\alpha}{\sqrt{\alpha^2-1}} - \frac{1}{\alpha}\right)$
are functions that determine the effect geometry has on the force.
We provide a fluid derivation of this system of equations
in Section XXX of the supplemental.
Similar envelope models are well established in the literature\cite{Lund:2014_emittance_def}; in fact,
the philosophy for presenting this model here is to have minimal complexity
for understanding the physics while
maintaining the agreement between the model and the dynamics of the
situation.

We note that Eq. (\ref{eq:wphi osc}) is not typically reported as part of the envelope equations;
however, equivalent forms of this expression 
can be found in the literature describing the equilibrium distribution of
SCPs (for instance, Eq. (5.36) of\cite{Reiser:1994_book}).
This equation is always derived from a consideration of the conservation of 
angular momentum, as we have done in Section XXX of the supplement.
However, our use of this equation is somewhat different from the standard use.
In most cases where this equation may be found, the researcher is looking at
a plasma at equilibrium.
Therefore, in addition to Eq. (\ref{eq:wphi osc}), equilibrium analyses 
also assume $R$ is constant in time, i.e. $\eta_R = 0$ and $\eta_Z = 0$
or more generally $(\eta_X,\eta_Y,\eta_Z) = (0,0,0)$. 
Substantial effort has been made to describe such distributions of 
interacting SCPs using these constraints [refs];
namely, we know ``equilibrium beams" satisfies Eq. (\ref{eq:wphi osc})
and $R =$ a constant
by adopting a non-uniform profile when the generalized 
angular momentum
is non-zero\cite{Bogema:1970_rigid_rotor}.
However, this well-established observation need 
not necessarily be true for far-from-equilibrium 
SCPs and beams.
Namely, if the radius is not constant, 
the distribution may be undergoing oscillations while remaining nearly uniform.
We examine this possibly controversial statement further in the next section.

For our first representative simulation, we initialized $10,000$ 
electrons with spatial coordinates sampled from a spherical-uniform distribution
with radius $2~\mu$m and evolved this distribution from rest under the influence of the particles' pair-wise 
repulsion and an external magnetic field strength of $4$ T.
Fig. \ref{fig:sphere:B} shows the evolution of the longitudinal magnetic moment
from both the $N$-particle simulation
and the envelope equations starting from rest. 
As can be seen, the
$N$-particle results and envelope prediction are in near perfect agreement.
As the longitudinal magnetic moment reverses direction twice during every oscillation,
the magnetic moment is aligned with the external magnetic field, i.e. is diamagnetic,
and opposite to the external magnetic field, i.e. is paramagnetic, for periodic 
intervals during this oscillation.
We note this oscillation is consistent with the Bohr-Van Leeuwen theorem\cite{Van:1921_theorem} as 
our system is far-from-equilibrium; likewise, it is not inconsistent with the standard attribution of 
magnetically confined plasmas as being diamagnetic\cite{ONeil:1995_overview} as such a description again assumes equilibrium-like 
conditions.

\begin{figure*}
  \centering
  \begin{tabular}{ccc}
    \subfloat[]{\includegraphics[width=0.3\textwidth]{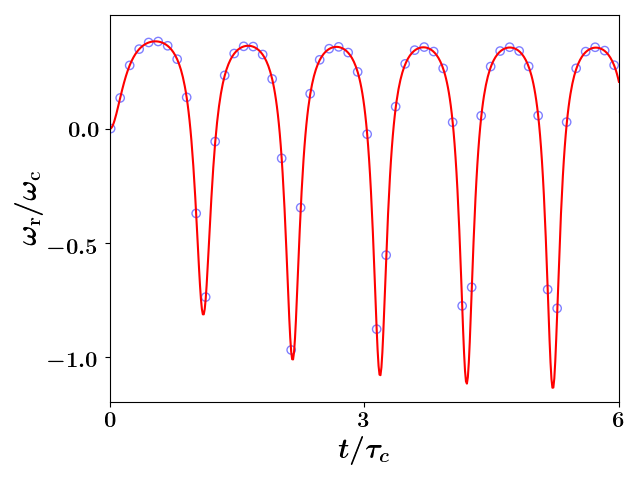}\label{fig:sphere:ang}}&
    \subfloat[]{\includegraphics[width=0.3\textwidth]{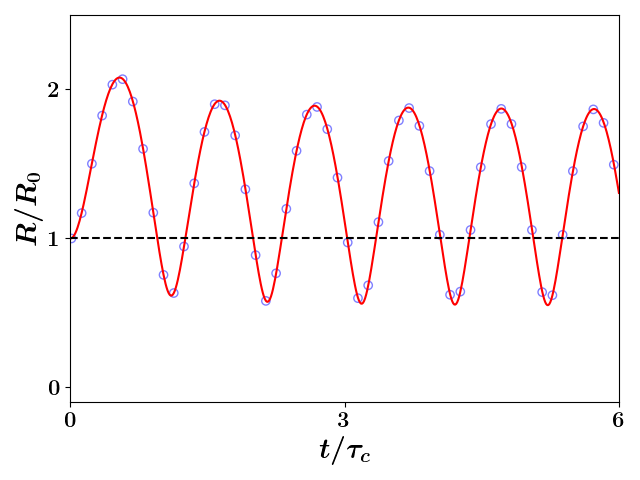}\label{fig:sphere:R}}&
    \subfloat[]{\includegraphics[width=0.3\textwidth]{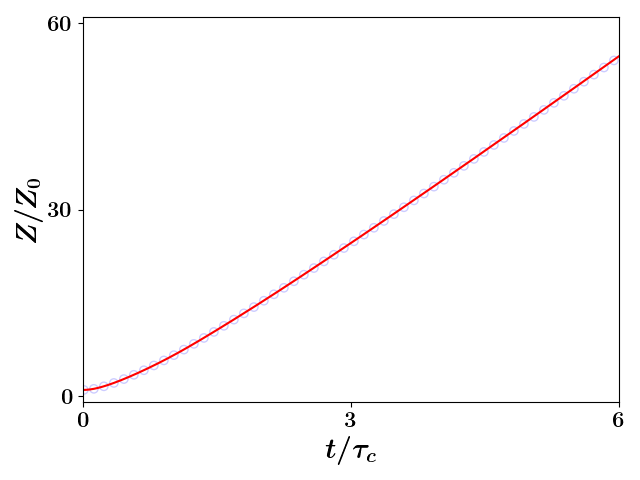}\label{fig:sphere:Z}}
  \end{tabular}
  \caption{
  Envelope statistics associated with simulation shown in Fig. \ref{fig:sphere:B}: (a) angular chirp,
  (b) transverse radius, and (c) longitudinal radius.  In all plots
  solid red lines predict the evolution of the statistic from envelope equations and 
  blue circles show the statistics calculated periodically during the simulation.
  In (b), The dashed black line is at $1$ indicates the initial radius of the SCP.
  The parameter $\tau_{\rm{c}}$ is the cyclotron period determined by the external magnetic field.
  }
\end{figure*}

As the longitudinal magnetic moment, $m_z$, in our notation is given by
\begin{align}
  m_z &= -\frac{e}{2} \omega_\text{r} R^2 \hat z\label{eq:mag mom},
\end{align}
we see that the sign of the magnetic moment of the SCP is determined by the direction of rotation
captured by the angular chirp.
The oscillation of the angular chirp for our first characteristic simulation
is displayed in Fig. \ref{fig:sphere:ang}; again, these results are in excellent agreement.
As the radial motion is captured in the model through the evolution of the parameter $R$, 
which is the sole time-dependent parameter in Eq. (\ref{eq:wphi osc}), 
we see that $\omega_\text{r}$ couples to the radial dynamics so that $\omega_\text{r}$ and $R$ oscillate at the same frequency.
The envelope predicted and $N$-particle simulated oscillation of $R$ and the unbound expansion of 
$Z$ are seen in Figs. \ref{fig:sphere:R} and \ref{fig:sphere:Z}, respectively.
Notice ${\tilde{\omega}}_{\text{r}}$
is negative for any value of $R$ that is less than $R_0$
for simulations starting from rest;
that is, the transition between the diamagnetic and paramagnetic states occurs
when the transverse radius crosses its original size as can be seen by comparing 
Figs. \ref{fig:sphere:B} or \ref{fig:sphere:ang} with Fig. \ref{fig:sphere:R}.
We note that the evolution of $Z$ seen in Fig. \ref{fig:sphere:Z} is typical of unconfined Coulomb explosion; namely,
the rapid acceleration of the growth in the statistic followed by a period of almost linear growth
leading to a consistent decrease in the density and hence a weakening of space charge effects.

\begin{figure*}
  \centering
  \begin{tabular}{cc}
    \subfloat[]{\includegraphics[width=0.4\textwidth]{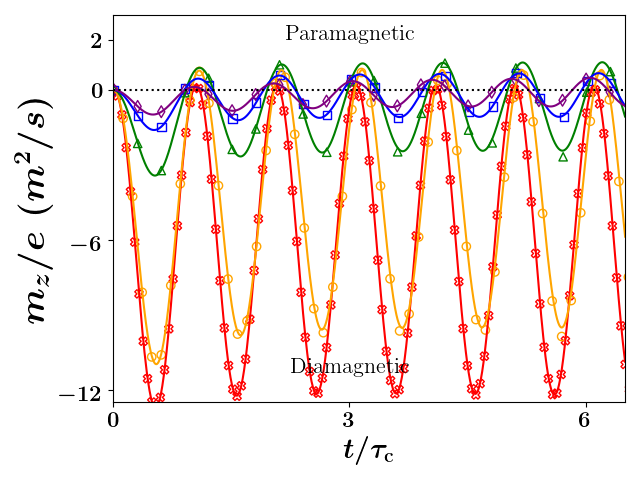}\label{fig:B_eff}}&
    \subfloat[]{\includegraphics[width=0.4\textwidth]{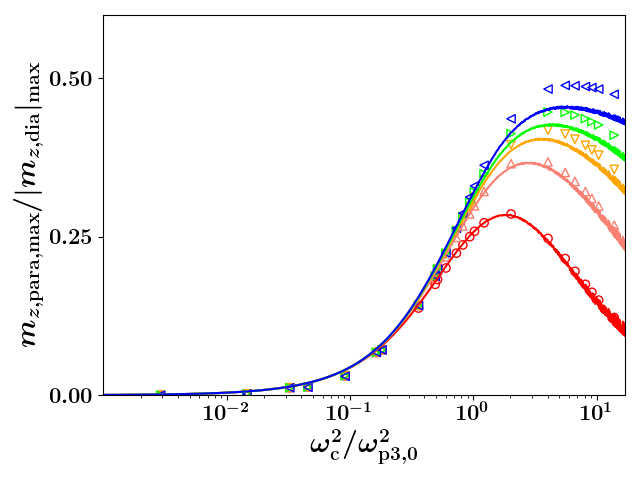}\label{fig:B_eff:ratio}}
  \end{tabular}
  \caption{
  The relative size of the paramagnetic oscillation is controlled by the magnetic field and density.
  (a) Simulated (scatter points) and envelope predicted (solid lines) evolution for the magnetic moment
  with $10,000$ electrons initially distributed uniformly within a sphere
  starting from rest.  
  Colors and markers of the plots indicate the value of $\frac{1}{{\tilde{\omega}}_{\rm{p}3,0}^2}$
  obtained from varying the radius and magnetic field: 
  red x's ($B = 1.0$T, $R=0.89~\mu$m, $\frac{1}{{\tilde{\omega}}_{\rm{p}3,0}^2} = 0.033$),
  orange circles ($B = 0.2$T, $R=4.5~\mu$m, $\frac{1}{{\tilde{\omega}}_{\rm{p}3,0}^2} = 0.16$), 
  green triangles ($B = 1.0$T, $R=4.5~\mu$m, $\frac{1}{{\tilde{\omega}}_{\rm{p}3,0}^2} = 1.0$),
  blue squares ($B = 2$T, $R=2.2~\mu$m, $\frac{1}{{\tilde{\omega}}_{\rm{p}3,0}^2} = 2.0$),
  purple diamonds($B = 1$T, $R= 6.7 ~\mu$m, $\frac{1}{{\tilde{\omega}}_{\rm{p}3,0}^2} = 14$).
  (b)
  Simulated (scatter points) and envelope predicted (solid lines) for the ratio of the paramagnetic to diamagnetic
  peak values as functions of $\frac{1}{{\tilde{\omega}}_{\rm{p}3,0}^2}$ visualized on a log scale.
  The statistic is shown 
  with different colors when it is calculated over different periods of time as the statistic evolves
  in time: red circles ($[0,1.5 \tau_{\rm{c}}]$),
  orange triangles ($[0,2.5 \tau_{\rm{c}}]$), green squares ($[0,3.5 \tau_{\rm{c}}]$),
  cyan stars ($[0,4.5 \tau_{\rm{c}}]$), and blue diamonds ($[0,6.5 \tau_{\rm{c}}]$).
  The $[0,5.5 \tau_{\rm{c}}]$ interval was skipped for presentation purposes as the curves converge toward an asymptote.
  }
\end{figure*}

As this effect is in $3D$ and the expanding longitudinal dimension results in a highly dynamic density, it 
is not apparent that this effect is the same for every magnetic field. 
As the only
frequencies relative to this situation are the cyclotron and plasma frequencies,
we consider this effect as a function of $\frac{1}{{\tilde{\omega}}_{\rm{p}3,0}^2}$.
In Fig. \ref{fig:B_eff}, we show how the 
oscillatory dynamics respond to changes in $\frac{1}{{\tilde{\omega}}_{\rm{p}3,0}^2}$ 
when all other parameters are held constant.
The amplitude of the oscillations decreases as the ratio between the frequencies is increased;
interestingly, though, the size of the oscillation's amplitude within the paramagnetic 
regime increases to a point at which further increasing of the magnetic field likewise begins to decrease
the size of the oscillation in the paramagnetic region.
In other words, at very low values of $\frac{1}{{\tilde{\omega}}_{\rm{p}3,0}^2}$, the self-magnetic field oscillation does
not enter very far into the paramagnetic region and such oscillation is effectively
turned on by increasing the magnetic field.  

To obtain better intuition into this effect, we quantified the relative size of the paramagnetic and diamagnetic 
oscillations by looking at the ratio of the minimum to the maximum self-magnetic moments, 
i.e. $\frac{m_{z,\rm{para,max}}}{|m_{z,\rm{dia}}|_{\rm{max}}}$.  
As can be seen in Fig. \ref{fig:B_eff}, the fist diamagnetic peak is somewhat larger than subsequent peaks, so this ratio
in our envelope treatment effectively uses the size of this first peak in the diamagnetic 
region to gauge the size of the subsequent peaks in the paramagnetic region.
Moreover, the subsequent peaks within the paramagnetic region slowly increase toward an asymptote
as can be seen in Fig. \ref{fig:B_eff:ratio} --- this is due to the SCP becoming less and less 
dense as it expands longitudinally resulting in the space-charge effect becoming less and less important in
determining where the oscillating SCP will reverse its inward motion.
On the other hand, as we increase the magnetic field, the number of oscillations
occuring before the space-charge effect becomes negligible increases. 
As we are
comparing simulation results at the same multiple of the the cyclotron period,
this means that
it takes the space-charge effect will emerge later in simulation with relative larger cyclotron frequencies.
This is the origin of the peak seen in Fig. \ref{fig:B_eff}.

As can be seen, increasing the ratio $\frac{1}{{\tilde{\omega}}_{\rm{p}3,0}^2}$ 
does appear to turn this oscillation on --- specifically, the turning on occurs when 
$\omega_{\rm{c}}^2/\omega_{\rm{p}3,0}^2$ is in the range $[0.1,3]$.  
Note that the ``Brillouin point" occurs when the initial
angular motion is rotating at $\frac{1}{2} \omega_{\rm{c}}$ and $\omega_{\rm{c}}^2/\omega_{\rm{p}3,0}^2 = 2$,
but here there is no initial angular motion and the oscillation is driven entirely by the initial expansion
of the electron SCP; therefore, the occurence of $2$ in this range is unrelated to the Brillouin point.

Comparing the results of the RMS envelope equations to the $N$-particle simulations, 
the general trends are captured by the model in both Figs. \ref{fig:B_eff} and \ref{fig:B_eff:ratio} 
and are almost exact at earlier times.
Notice that the deviations in the ratio monotonically increase during the simulation and are larger for 
stronger external magnetic fields, 
and a close inspection of the magnetic moment oscillation for higher magenetic fields can be seen in Fig. \ref{fig:B_eff}.
Coincidentally, these are the cases that have larger emittance growth, so we infer
these deviations are due to accumulated emittance growth that  
are not included in Eqs. (\ref{eq:3D ddot}). 
We will model the effect of RMS emittance in this situation in a later section, but such
treatment will still consider the emittance as being conserved --- just non-zero --- but
still not dynamic as it is in the simulations.


\section{Rigid-rotor non-equilibrium and emittance-induced shear}

As mentioned in the previous section, substantial work has been done to describe the
steady-state properties of confined plasmas[refs].
Initial work by Brillouin described a uniform steady-state
where the angular chirp is given by the Larmor frequency, 
$\omega_{\rm{r}} = \omega_{\rm{L}} = \frac{1}{2} \omega_{\rm{c}}$\cite{Brillouin:1945_SCP_mag_field},
and subsequent work has extended this understanding
to non-uniform steady-state distributions rotating with angular chirps in the range
$[0,\omega_{c}]$ known as rigid-rotor equilibrium.

To our knowledge, there is no clear quantitative definition of rigid-rotor equilibrium;
so we develop one here.  
If we were to calculate the angular kinetic energy of a rigid-rotor,
we would expect that the amount of angular kinetic energy 
predicted by the linear chirp, i.e. $\frac{N}{10} m \omega_{\rm{r}}^2 R^2$,
should be much larger than the amount of angular kinetic energy associated
with random motions.
Consider the
angular velocity of the $i^{th}$ particle, $v_{\phi,i}$.
We know that the residual for the $i^{th}$ particle between the chirp-predicted and
measured angular velocity can be written as
\begin{align}
  \delta_{v_{\phi},i} &= v_{\phi,i} - \omega_{\rm{r}} r_i.
\end{align}  
where $r_i$ and $v_{\phi,i}$ are the radial and angular velocity coordinates for particle $i$.
Therefore the variance of the residuals can be written as
\begin{align}
  s_{\delta_{v_{\phi}}}^2 &= s_{v_{\phi}^2} + \omega_{\rm{r}}^2 s_r^2  - 2 \omega_{\rm{r}} s_{v_{\phi} r}.\label{eq:var resid vphi}
\end{align}
Note that $s_r \ne R$ as $r > 0$ for all $r$ whereas $\bar{x}$ and $\bar{y}$ are approximately $0$.
As  $\frac{N}{2} m s_{\delta_{v_{\phi}}}^2$
represents the portion of the kinetic energy that is associated with motions that are not linearly correlated
and is often attributed to being random, it follows
that the ratio
\begin{align}
  \nu &= \frac{1}{5} \frac{\omega_{\rm{r}}^2 R^2}{s_{v_{\phi}^2} + \omega_{\rm{r}}^2 s_r^2  - 2 \omega_{\rm{r}} s_{v_{\phi} r}}\label{eq:rrstat}
\end{align}
can quantify the rigid-rotor approximation; we call $\nu$ the rigid-rotor statistic.
Indeed if the rigid-rotor statistic 
is larger than $1$, the majority of the angular kinetic energy is associated with the motion described by the angular chirp. 

\begin{figure*}
  \centering
  \begin{tabular}{ccc}
    \subfloat[]{\includegraphics[width=0.3\textwidth]{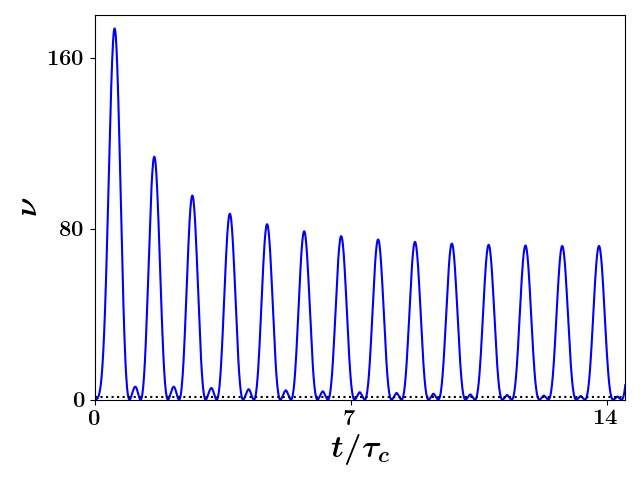}\label{fig:rrstat:full}}&
    \subfloat[]{\includegraphics[width=0.3\textwidth]{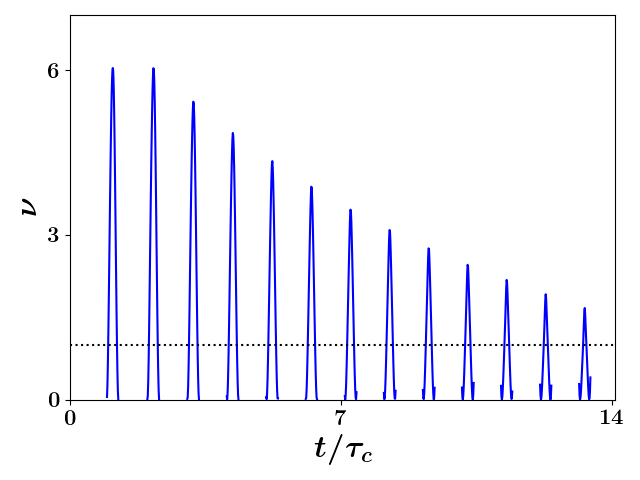}\label{fig:rrstat:below}}&
    \subfloat[]{\includegraphics[width=0.3\textwidth]{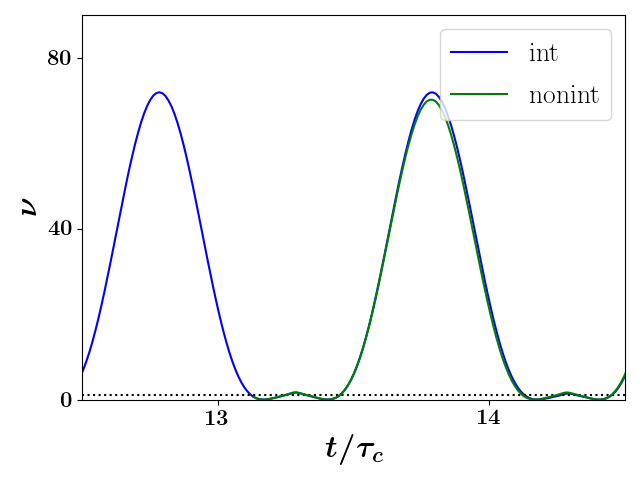}\label{fig:rrstat:non}}
  \end{tabular}
  \caption{
  Evolution of the rigid-rotor statistic, $\nu$, that is the ratio of the angular kinetic energy associated with
  the angular chirp to the kinetic energy of all angular motion less the chirped angular motion kinetic energy.
  (a) Evolution of this statistic across the full interacting simulation.
  (b) Evolution of this statistic restricted to the periods of time when $R < R_0$ for the interacting simulation.
  (c) Evolution of this statistic for the final oscillations of the non-interacting (green line) and interacting (blue line) simulations.
  }
\end{figure*}

For a second characteristic simulation, we show the rigid-rotor statistic as a function
of time in Fig. \ref{fig:rrstat:full}.
This second simulation again had $10,000$ electrons but was initially sampled from the 
spherical uniform distribution with initial radius $R_0 = 5~\mu$m from rest with a magnetic field strength of 
$1$ T.
As can be seen in this figure, the rigid-rotor statistic is much larger than $1$ except when it crosses
$R_0$ at which time the SCP is in the process of reversing its direction of rotation; at this time, the angular chirp is zero and therefore
the kinetic energy associated with the coherent portion of the rotation captured by this statistic is zero as well. 
Such dynamics should not be described as rigid-rotor-like.  
However, as the radius continues to decrease,
the angular chirp becomes negative and the rigid-rotor statistic once again become greater than $1$.
This means the distribution begins to rigidly rotate in the opposite direction after briefly passing through a state
dominated by random angular motion.
Due partially to the fact that the expansion beyond $R_0$ is relatively much larger than the contraction within $R_0$,
the rigid-rotor statistic for this latter region does not get as large.  
Moreover, the rate at which the rigid rotor statistic  evolves within the larger than $R_0$ and smaller than $R_0$
regions differ as can be seen in Figs. \ref{fig:rrstat:full} and \ref{fig:rrstat:below}.

It is unclear from $N$-particle simulations alone to what extent each of 
non-zero emittance, space-charge, and emittance evolution effects contribute to
evolution of the rigid-rotor statitics seen in Fig. \ref{fig:rrstat:full}
especially once the SCP has expanded significantly.
To obtain insight into such dynamics, we re-simulated the last two
transverse oscillations without inter-particle forces
starting with the same configuration produced from the interacting simulation
of Coulomb explosion.
We show the comparison between the results of these non-interacting and interacting simulations 
in Fig. \ref{fig:rrstat:non}.
We note that the evolution of the rigid-rotor statitic is very
similar in both cases suggesting
that non-interacting-like behavior dominates towards the end of the
transverse oscillations; such behavior would also suggest that emittance
growth is responsible for the small deviations.
On the other hand, we would expect that emittance effects should dampen the
rigid-rotor statistic, but what is seen is the opposite --- that
is, the rigid-rotor statistic for the interacting simulation
is larger than the non-interacting simulation.

\begin{figure*}
  \centering
  \begin{tabular}{ccc}
    \subfloat[]{\includegraphics[width=0.3\textwidth]{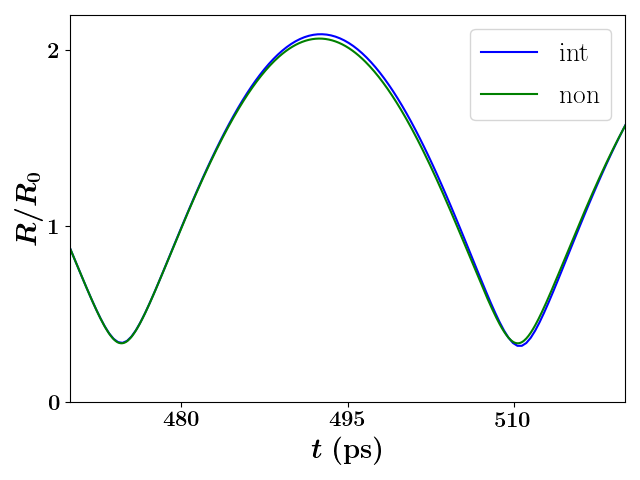}\label{fig:comp:sx}}&
    \subfloat[]{\includegraphics[width=0.3\textwidth]{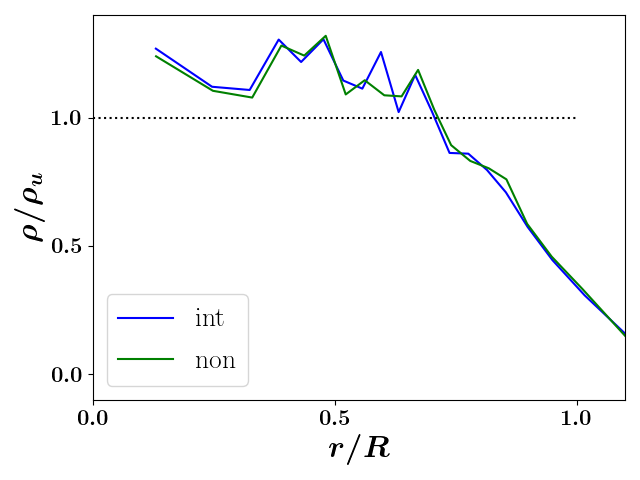}\label{fig:comp:den}}&
    \subfloat[]{\includegraphics[width=0.3\textwidth]{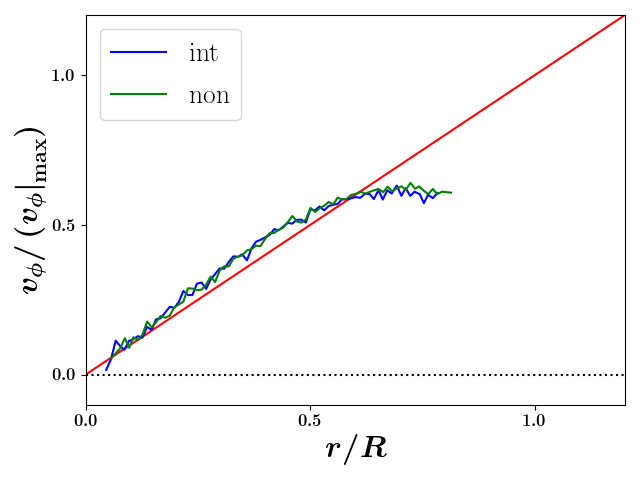}\label{fig:comp:phase}}
  \end{tabular}
  \caption{
  Comparison between radial evolution and properties of the phase space distribution 
  toward the end of the interaction simulation (blue lines) 
  and non-interacting simulation (green line) --- the non-interacting simulation
  was initialized with the configuration of the interacting simulation when
  the simulated transverse radius crossed $R_0$ prior to the $13^{th}$ minimum.
  (a) Plot of the radial statistic as a function of time.
  (b) Plot of the normalized radial densities (each density is normalized by it own maximum $R$).
  Dotted black line shows the profile of the ideal uniform distribution used to initialize
  the interacting simulation.
  (c) Plot of the normalized angular velocity phase space.  The red-line shows the 
  expected chirp for an ideal uniform distribution.
  }
\end{figure*}

As the rigid-rotor statistic is correlated with the transverse radius of the SCP,
one explanation for this effect could be if the interacting simulation were to obtain a larger
radius than the non-interacting simulation thus affecting the numerator of the rigid-rotor statistic.
In Fig. \ref{fig:comp:sx}, the evolution of the non-interacting and interacting radii
are shown. 
As can be seen, the simulated SCP with no interaction predicts a less extreme
transverse oscillation.
Thus this explanation is consistent with the observed difference between the
maxima of the rigid-rotor statistics in Fig. \ref{fig:comp:phase}; however, a small deviation is also observed
at the minimum radius, and the rigid-rotor statistic does not display the complementary 
deviation in the rigid-rotor statistic during this time period.
So something more is going on near the transverse minima that
compensates for the radial effect seen at the maximum.


We explored the
densities predicted near the final simulated transverse minimum
to understand if a space-charge shift in the bunch profile could be compensating 
for the radial-induced expected deviation near the minima. 
As the second-order moments are sensitive to profile changes,
such a profile change should result in a change in the rigid-rotor statistic
that we would then attribute to the use of second order moments.
We show the transverse radial density at the $14^{th}$ transverse radial minimum for both
simulations in Fig. \ref{fig:comp:den}.
As can be seen, the simulated radial densities do in fact deviate substantially from the
theoretically uniform distribution, but they do not deviate from one another.
Specifically, a clear edge effect that is well discussed in the equilibrium literature [refs]
arises when the bunch reaches its minimum transverse radius, but again, this 
occurs in both the interacting and non-interacting simulations.
Therefore, this density effect cannot be causing the difference in the rigid-rotor statistic.
Further, as the only portion of the non-interacting dynamics that is driving distribution change
is the emittance, we conclude that the edge effect is solely an emittance effect in both cases.

A second possible mode of compensation would be if our
statistical estimate of the kinetic energy in the angular mode 
from the angular chirp 
could differ between the two simulations if non-linearities 
were to arise. 
We show the angular phase-space at 
the $14^{th}$ transverse radial minimum for both
simulations in Fig. \ref{fig:comp:phase}.
As in the radial density, a clear edge effect, which could be described as a shear, can be seen; however,
again the same effect is seen in both the interacting and non-interacting simulations.
Therefore, we conclude that the compensation in the rigid-rotor statistic is not driven by 
this shear effect and further that the shear is likewise a result of emittance. 

\begin{figure}
  \centering
  \begin{tabular}{c}
  \subfloat[]{\includegraphics[width=0.35\textwidth]{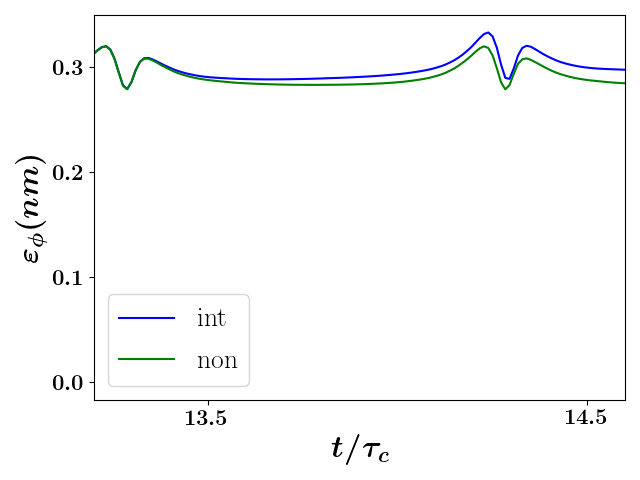}\label{fig:ang em}}\\
  \subfloat[]{\includegraphics[width=0.35\textwidth]{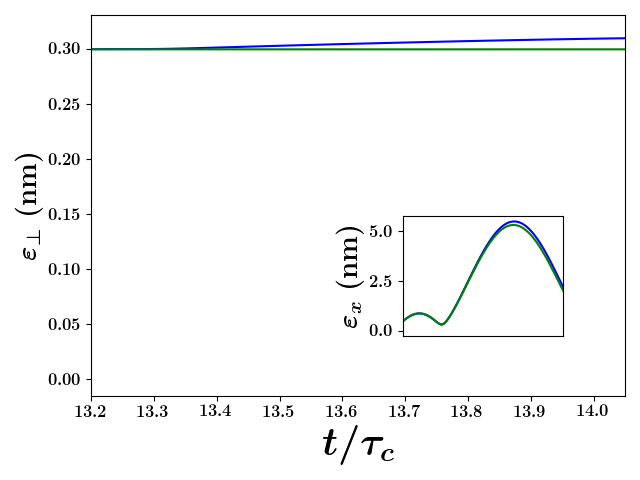}\label{fig:therm em}}
  \end{tabular}
  \caption{
  Evolution of the emittances toward the end of the non-interacting (green lines)
  and interacting (blue lines) simulations.  
  (a) Evolution of the angular emittance defined through Eqs. (\ref{eq:ang em}) and (\ref{eq:var resid vphi}).
  (b) Evolution of the cylindrically-symmetric thermal emittance defined by Lund \textit{et al}.
  The evolution of the standard definition of the $x$-emittance over the same time period is shown in the inset;
  the large variation seen in this measure makes it inappropriate to use and
  is due to the transverse coupled dynamics.
  }
\end{figure}

A third possible compensatory mechanism is if the denominator
in Eq. (\ref{eq:rrstat}) were to somehow differ between the two simulations.
To examine this denominator, we point out that normalized RMS emittance,
standardly defined [refs] as 
\begin{align}
  \varepsilon_{x} &= \sqrt{s_x^2 s_{v_{x}}^2 - s_{x,v_x}^2}
\end{align}
and defined transversely in the presence of cylindrical-symmetry about a magnetic field by\cite{Lund:2014_emittance_def}
\begin{align}
  \varepsilon_{\perp} &=  \sqrt{ \frac{1}{4}(s_{v_{x}}^2+s_{v_y})^2 - R^2(\eta_R^2 + \omega_r^2)}\label{eq:Lund},
\end{align}
is effectively the velocity spread times the spatial spread when the bunch is
projected to a single dimension [refs].
Therefore, we multiplied the velocity spread, or the square root of the residual variance
from Eq. (\ref{eq:var resid vphi}),
by $R$ to define an ``angular emittance",
\begin{align}
  \varepsilon_{\phi} &= R s_{\delta_{v_{\phi}}}\label{eq:ang em}.
\end{align}
As can be seen in Fig. \ref{fig:ang em}, this measure is very roughly
conserved during the non-interacting simulation.
Interestingly, it is also roughly the same value as the cylindrically-symmetric emittance
seen in Fig. \ref{fig:therm em}.
The small emittance growth seen in both the angular and transverse emittances
appears to contribute to  the fact that the rigid-rotor statistic is smaller
at the minimum of the transverse width than should be expected from the evolution of the radius
alone.

We note that this roughly conserved angular emittance has not been described in the literature
but can be argued to be the complement of the round-beam thermal emittance defined by Lund \text{et al}\cite{Lund:2014_emittance_def}
presented in Eq. (\ref{eq:Lund}).
Also interestingly, obtaining the Cartestian covariances
and re-sampling the distribution to make an equivalent
initial condition [not shown] neither captures the correct angular emittance
nor the evolution of the rigid-rotor statistic.
This suggests that the standard statistics are not capturing all of the spatial correlations
present in the SCP.

\section{Effect of non-zero emittance on magnetic state reversal in simulations}
We now consider cases where the SCP starts with some temperature, which we quantify by the 
dimensionless emittances
defined by
\begin{subequations}
\begin{align}
  \varepsilon_{\perp}^2 &= \frac{R^2}{R_0^4 \omega_{\rm{c}}^2} \left(\frac{s_{v_x}^2 + s_{v_y}^2}{2} - \left(\eta_R^2 + \omega_{\rm{r}}^2 \right)R^2\right)\\
  \varepsilon_{z}^2 &= \frac{1}{R_0^4\omega_{\rm{c}}^2}\left(s_z^2s_{v_z}^2 - s_{z,v_z}^2\right)
\end{align}
\end{subequations}
where $\varepsilon_{\perp}$ is a dimensionless version of Eq. (\ref{eq:Lund}).
Following the formulation of the envelope equations by Lund \textit{et al}\cite{Lund:2014_emittance_def}, we add standard definitions
of emittance to our equations with the additional assumption that we are starting with
non-zero angular motion, $\omega_{\rm{r},0}$, resulting in changes to Eqs. (\ref{eq:3D omega_R}), (\ref{eq:wphi osc}), (\ref{eq:3D Cz}):
\begin{subequations}
\begin{align}
  \frac{d{\tilde{\eta}}_R}{d\tau} &= \frac{1}{2} {\tilde{\omega}}_\text{\rm{p}3,0}^2 \Gamma_1\left(\alpha\right) \frac{1}{\tilde{R}^2\tilde{Z}} - \frac{1}{4} \nonumber\\
   &\quad + \left(\left({\tilde{\omega}}_{\rm{r},0} - \frac{1}{2}\right)^2 + \varepsilon_{\perp}^2 {\tilde{\omega}}_\text{\rm{p}3,0}^2 \right) \frac{1}{\tilde{R}^4} - {\tilde{\eta}}_R^2\\
 {\tilde{\omega}}_{\text{r}} &= \frac{1}{2} - \left({\tilde{\omega}}_{\rm{r},0} - \frac{1}{2}\right)\frac{1}{{\tilde{R}}^2}\label{eq:w no dim}\\
  \frac{d{\tilde{\eta}}_Z}{d\tau} &= \tilde{{\omega}}_\text{\rm{p}3,0}^2 \Gamma_2\left(\alpha\right) \frac{1}{\tilde{R}^2\tilde{Z}} + \frac{\varepsilon_{z}^2}{\tilde{Z}^4} - {\tilde{\eta}}_Z^2
\end{align}
\end{subequations}
Our addition of the transverse and longitudinal emittance terms can be argued analogous to 
Lund \textit{et al.} or derived completely from a statistic standpoint (in preparation);
angular emittance is not added to Eq. (\ref{eq:w no dim}) as the equation is a statement of conservation of
angular momentum.
The addition of the dimensionless normalized RMS emittance terms captures the effect of motion that deviates from the
linear chirps, sometimes described as thermal effects [refs], that result in an outward-like force on the 
RMS envelope parameters.
This is analogous to temperature pressure in fluid models [refs], but is formulated from a discrete statistical vantage point
where no theoretical approximations are needed.

\begin{figure*}
  \centering
  \begin{tabular}{cc}
    \subfloat[]{\includegraphics[width=0.4\textwidth]{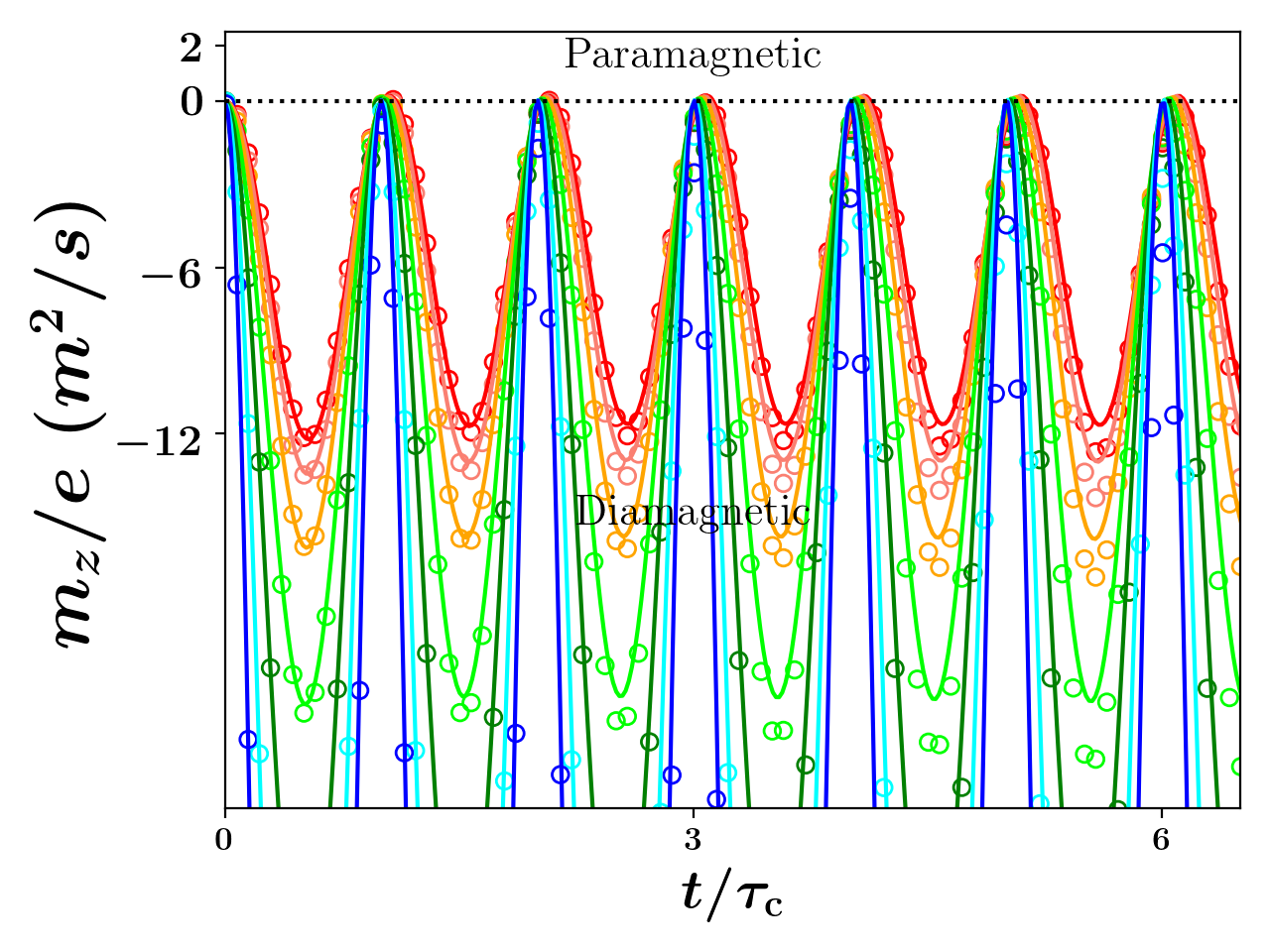}\label{fig:em:mult}}&
    \subfloat[]{\includegraphics[width=0.4\textwidth]{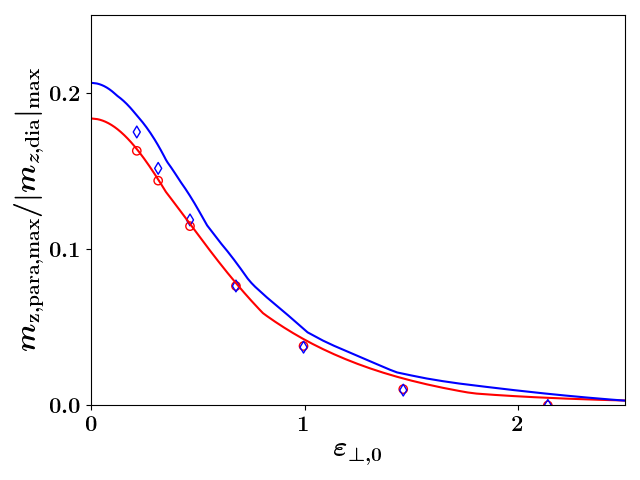}\label{fig:em:off}}
  \end{tabular}
  \caption{
  Increasing the emittance dampens the size of the paramagnetic field.
  (a)  Simulated (scatter points) and predicted (solid lines) evolution
  of the magnetic moment of an electron SCP where all initial conditions had the same 
  $\frac{1}{{\tilde{\omega}}_{\rm{p}3,0}^2} = 0.511$ with the same initial radius of
  $2.24 ~\mu$m
  but different values of $\varepsilon_{\perp,0}$ (denoted by different colors):
  blue circles ($\varepsilon_{\perp,0} = 0.21$),
  cyan triangles ($\varepsilon_{\perp,0} = 0.46$),
  green squares ($\varepsilon_{\perp,0} = 0.68$),
  orange diamonds ($\varepsilon_{\perp,0} = 1.0$),
  red stars ($\varepsilon_{\perp,0} = 2.1$),
  (b)
  Simulated (scatter points) and envelope predicted (solid lines) for the ratio of the paramagnetic to diamagnetic
  peak values as a function of $\frac{1}{{\tilde{\omega}}_{\rm{p}3,0}^2}$.
  As the ratio only evolves a little for these initial conditions, the statistic is shown 
  with two different colors when it is calculated over different periods of time: red circles ($[0,1.5 \tau_{\rm{c}}]$),
  blue diamonds ($[0,6.5 \tau_{\rm{c}}]$).
  }
\end{figure*}

To examine the effect of the emittance on the diamagnetic/paramagnetic oscillation, we conducted $N$-particle simulations of
initially spherically symmetric Coulomb explosion, initially with $\varepsilon_{z,0} = \varepsilon_{\perp,0}$, 
and integrated the envelope equations. 
For the initial conditions we held $\frac{1}{{\tilde{\omega}}_{\rm{p}3,0}^2}$ fixed
and varied $\varepsilon_{\perp,0}$.  The evolution of some such characteristic simulations can be seen in 
Fig. \ref{fig:em:mult}.
What is apparent is that increasing the initial emittance leads to larger oscillations, shifts the frequency
of oscillations higher and decreases the size and duration of the magnetic moment within the paramagnetic region ---
in a way, this is nearly the opposite of the effect of increasing the confining magnetic field.
A plot of the ratio of the maximum self-magnetic moments in each region with such initial conditions viewed as a function of $\varepsilon_{\perp,0}$
can be seen in Fig. \ref{fig:em:off}.  
The envelope data is shifted rightward relative to the $N$-particle simulated data as the $N$-particle
simulations undergo disorder induced heating that increases the emittance.
As can clearly be seen, increasing the dimensionless emittance beyond about 0.1
decreases the paramagnetic/diamagnetic statistic, and the effect is again turned off once the dimensionless emittance
is increased past about 2.  

\begin{figure}
  \includegraphics[width=0.4\textwidth]{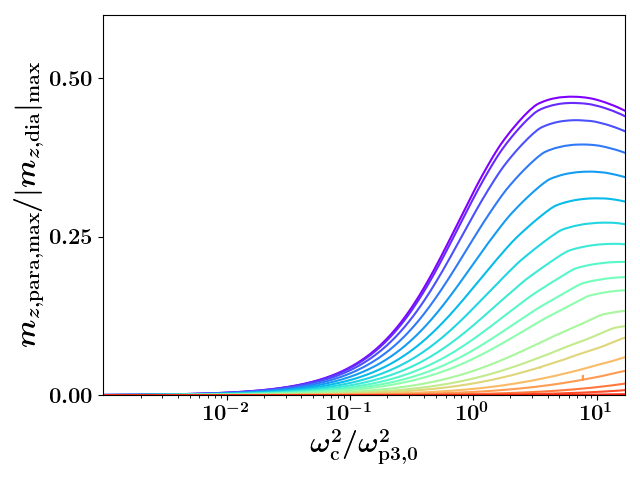}
  \caption{\label{fig:both}
  Emittance moderation of the diamagnetic/paramagnetic oscillation
  obtained from our model.
  The ratio between the most positive and absolute value of the most negative magnetic moment
  was determined within the period $[0,6.5 \tau_c]$, where $\tau_c$
  represents the cyclotron period.  Colors were chosen on a rainbow gradient from purple to red and corresponding
  to the 
  dimensionless emittances of 0, 0.1, 0.2, 0.3, 0.4, 0.5, 0.6, 0.7, 0.8, 0.9, 1, 1.2, 1.4, 1.6, 2, 2.5, 3.5, 
  5, and 10, respectively.
  }
\end{figure}

Our envelope model predicts the behavior of the $N$-particle simulation well especially before
the emittance has grown significantly. We have also produced a figure detailing the interaction between
$\frac{1}{{\tilde{\omega}}_{\rm{p}3,0}^2}$ and the emittance 
using just the envelope predictions within the first $6.5$ plasma periods.  
The competitive nature of these two effects can be seen in
Fig. \ref{fig:both};
namely, an increase in the emittance requires larger values of $\frac{1}{{\tilde{\omega}}_{\rm{p}3,0}^2}$ before
a substantial paramagentic state is observable.
Further, increasing the emittance delays at what ratio of $\frac{1}{{\tilde{\omega}}_{\rm{p}3,0}^2}$ this effect arises 
as can be seen by the fact that the curves with 
significant emittance in Fig. \ref{fig:both} appear to be scaled versions of the successive evolution of this statistic seen in
Fig. \ref{fig:B_eff:ratio}. 
As this delay enables additional effects, like cyclotron radiation and other electromagentic effects, 
outside of the scope of this paper
to occur and further drive the SCP toward equilibrium, 
sufficiently large emittance will effectively turn this effect off. 

So far, our analysis has been principally of numerical results.  
The spheroidal envelope model does not have an exact analytical solution; 
however, simpler models do allow more analysis.
If we were to match such a model to an experiment where the space charge effect 
remained important even once the bunch had become significantly oblate, it would 
require us to fit the initial parameters of the model to the 
parameters observed
analogous to our non-interacting treatment.
We analyze this continuous-beam model.

The continuous model is very similar to the spheroidal model.  The
principal differences are: (1) there is no longitudinal dimension to model
and (2) the geometric factor is absorbed into the plasma frequency
giving a cylindrically symmetric plasma frequency.
We denote this new plasma frequency by 
$\omega_{\text{p2,0}} = \sqrt{\frac{e^2}{m\epsilon_0} \frac{n_\text{z}}{\pi R_0^2}}$
where $n_z$ is the number density along the length of the beam.
The subsequent radial oscillation is determined by
\begin{align}
  \frac{d{\tilde{\eta}}_R}{d\tau} &= \frac{1}{2} {\tilde{\omega}}_{\text{p2,0}}^2 \frac{1}{\tilde{R}^2} - \frac{1}{4} + \left(\left({\tilde{\omega}}_{\text{r,0}} - \frac{1}{2}\right)^2 + \varepsilon_{\perp,0}^2 {\tilde{\omega}}_{\text{p2,0}}^4\right) \frac{1}{\tilde{R}^4} \label{eq:2D ddotR}.
\end{align}
Note the evolution of the angular frequency is only changed by the third dimension only
through its dependence on $R$ as we assume angular momentum is completely
along the axis of symmetry.
We note that many equivalent models to this can be found in the literature\cite{Dubin:1993_equilibrium,Reiser:1994_book}. 
While this model depends on three parameters, ${\tilde{\omega}}_{\text{p2,0}}^2$,
${\tilde{\omega}}_{\text{r,0}}$, and the initial
radial velocity, the freedom to assign the initial phase of the oscillation
allows us to set one of the initial velocities, by convention the initial radial velocity, to zero.
Thus the properties of the oscillations predicted by this model can be understood within the 
$\left({\tilde{\omega}}_{\text{r,0}},{\tilde{\omega}}_{\text{p2,0}}^2\right)$ 
parameter space when $\varepsilon_{\perp,0}$ is held fixed, and we will conduct our analysis within this
parameter space for various values of $\varepsilon_{\perp,0}$.

We note that this parameter space is
standard in descriptions of rigid-rotor equilibria;
namely, this parameter space is typically used to visualize
the parabola of zeros of the effective force
that describe the well known rigid-rotor equilibria\cite{Gould:1995_review}.
In the accelerator literature where continuous beam behavior near rigid-rotor states is discussed,
small radial oscillations about a rigid-rotor equilibrium are 
described as RMS-mismatched induced transverse ripples along the longitudinal coordinate\cite{Reiser:1994_book};
however, such descriptions focus solely on the effect of $R$ and
not its coupling to $\omega_\text{r}$, which we emphasize as being important here.

\begin{figure}
  \includegraphics[width=0.4\textwidth]{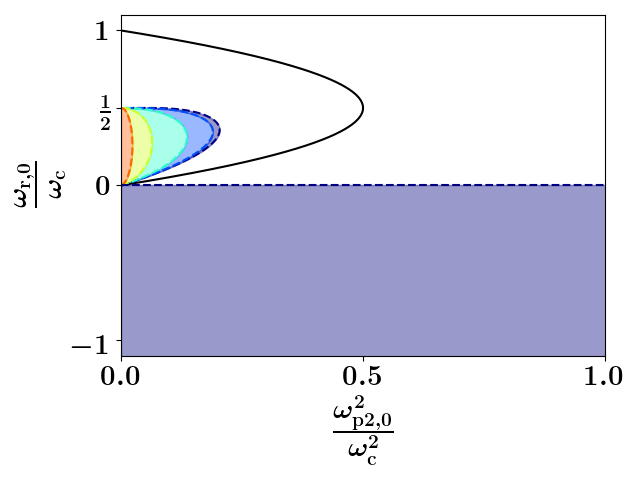}
  \caption{\label{fig:diag:pd}
  Diagram
  indicating the regions (shaded blue) of $\left({\tilde{\omega}}_{\text{p2,0}}^2,{\tilde{\omega}}_{\text{r,0}}\right)$ parameter space
  where the diamagnetic/paramagnetic phase transition occurs in 
  continuous-beams. The solid black parabola represents 
  the rigid-rotor equilibria in this space. 
  }
\end{figure}

A diagram denoting all of the regions
of parameter space  where the initial conditions of the continuous-beam model will experience the phase
transition for various values of $\varepsilon_{\perp,0}$ is displayed in Fig. \ref{fig:diag:pd}.
One interesting aspect of this diagram is the fact that the two shaded regions, one inside the rigid-rotor parabola
and a second describing the half-plane with $\omega_\text{r,0} < 0$, describe the same oscillations.
That is, our convention of no initial radial velocity is ambiguous as there exists 
two times per period with zero radial velocity: (1) the time where the radius is maximum 
(corresponding to the entire region of parameter space inside the 
rigid-rotor parabola)
and (2) the time at
which the radius is minimum (that region's complement).  
Deducing the fact that all initial conditions within the half plane where $\omega_\text{r,0} < 0$ 
produce the paramagnetic oscillations is simple;
we know that the point in the oscillation that corresponds to the maximum radius
is inside the parabola and has $\omega_\text{r} > 0$, so all such initial states in this region
must change their direction of rotation at some time.  
The initial states within the rigid-rotor equilibrium parabola that experience the phase transition 
requires more analysis, and we present the mathematical description of this region's boundary
in Section VI of the supplement.

\section{Discussion}

We have described the dynamics of Coulomb explosion of an electron SCP within a magnetic field
where we have found that the rotation of the SCP can spontaneously reverse direction
due to a combination of the space-charge effect and the conservation of angular momentum.
This reversal of the rotational direction of the SCP 
results in the SCP periodically switching between diamagnetic and paramagnetic 
states.  
This observation is surprising given the established diamagnetic nature of 
SCPs in ubiquitous Brillouin flow models.  
We showed that this behavior is turned on by increasing the magnetic field, and turned off
by increasing the emittance or thermal kinetic energy in the SCP.
Moreover, we showed that this behavior arises from coupling between the 
radial and angular motion introduced by the magnetic field, and
presented equations that describe this coupling.
We modeled the radial and hence angular motion of such SCPs with 
mean-field particle-particle interactions through envelope equations and showed that the
envelope equations captures the angular dynamics to the same extent that the radial dynamics are captured.

It is clear that SCPs particularly in traps rapidly evolve toward an equilibrium distribution.
During this time, the plasma typically undergoes disorder induced heating\cite{Murillo:2006_dih}
as well as cooling through cyclotron radiation\cite{ONeil:1980_cooling}.
However, a full kinetic picture of how such a plasma evolves from a possibly
far-from-equilibrium initial state with substantial inter-particle correlation 
to the eventual equilibrium distribution 
where particle inter-particle correlations are treated randomly
is not apparent within the literature.  
Our Coulomb explosion model provides insight into some possible early dynamics that
could further evolve the inter-particle correlations if relativistic effects were 
to be properly considered in future work.

We note the parameters we use have statistical definitions.
We have been calling the description of the evolution of such
parameters statistical kinematics although they are sometimes known as
moment equations in the literature\cite{Sacherer:1971_envelope}.
These statistical definitions have the benefit of being applicable to
arbitrary distributions; in this work, this is true for the radial measure, $R$, 
angular measure, $\omega_{\rm{r}}$, and magnetic moment, $m_z$.  
However, to capture the time dependence of one of these parameters, specifically the radial measure, 
we needed to apply mean field theory using the uniform distribution to avoid space-charge non-linearities
for which general techniques have yet to been developed.
So despite the fact that Eq. (\ref{eq:wphi osc}) relates the dependence of $\omega_{\rm{r}}$ on
$R$ in any generic distribution, 
we can only capture the time evolution through the use of
the uniform distribution
for which an analytic expression for the space-charge force is available.

The insight into the radial/angular coupling adjusts the 
typical depiction of the SCP breathing mode as a periodic inward and 
outward motion\cite{Dubin:1996_normal_modes}.  
Instead, we've shown that the angular motion oscillates coherently with the radial motion.
That is, instead of moving in and out, the SCP also rotates with rate of the rotation specified
by the size of the radius.
The microscopic origin of this is of course the cyclotron motion of the particles,
and we have provided a simplified schematic of this motion for the non-interacting distribution 
in the supplemental documentation.

The fundamental theory can be tested experimentally by careful accounting 
for the mechanical angular momentum
of a beam.
We note that generalized Courant-Snyder theory\cite{Qin:2013_gen_courant}
provides tools to track the rotation angle of the envelope matrix in the presence of 
transverse-coupled dynamics, and this rotation can alternatively be conceptualized as the angular momentum
of the beam as we've done here.
Therefore, that theory and the tools we present here may 
be used to do such accounting.
More specifically,
a long solenoid can be used to introduce mechanical angular motion to a prolate
SCP in an accelerator setting; the amount of angular momentum can be determined using 
Eq. (\ref{eq:wphi osc}) and (\ref{eq:2D ddotR}) and knowledge of the
angular momentum kick across the solenoid's fringe.  
As the SCP evolves in a region without a magnetic field, 
these equations may again be employed with
$\omega_\text{L} = 0$ and the $\omega_{r,0}$ introduced from the solenoid.
The SCP could then be passed through a second long solenoid.
By tuning the length of the first solenoid, the magnetic field strengths, and the separation
distance between the solenoids for an SCP of a known length and number of electrons, 
it should be possible to induce a near equilibrium SCP once it enters the 
second solenoid. 

\section{Methods}
\subsection{Analytic models}
We derived the analytic model both from a fluid and from a statistical perspective.
Analytic results were compared in order to confirm the accuracy
of the derivation and our understanding of the physics.
Details of the analytic fluid derivation of the model are provided in the supplement.

\subsection{Numerical simulations}

We conducted $N$-particle simulations and numerically integrated the 
envelope equations (Eqs. (\ref{eq:2D ddotR}) and (\ref{eq:3D ddot})). 

For the $N$-particle simulations, spatial coordinates were sampled from the specified initial distribution
and propagated using Velocity Verlet algorithm\cite{Verlet:1967_algorithm} with the self field determined 
by the Fast Multipole Method algorithm\cite{Gimbutas:2015_fmmlib3d} 
or set to zero (for the non-interacting simulations).
Time steps were set to between $1/300$ and $1/1000$ $\omega_{\rm{c}}$.
Parameters included:
\begin{enumerate}
  \item{Fig. 1: 
  We simulated (non-interacting)
  $10k$ electrons initially uniformly, spherically-symmetrically distributed 
  within the radius $R_0 = 10 ~\mu$m 
  with rotation rate $\omega_{\text{r},0}  = 0.25 \omega_\text{c}$ 
  about the axis of an external magnetic field strength of $1$ T.}
  \item{Fig. 2: 
  We simulated (non-interacting)
  $10k$ electrons initially Gaussian, spherically-symmetrically distributed 
  with standard deviation $ \sigma_r = 10 ~\mu$m 
  with rotation rate $\omega_{\text{r},0}  = 0.25 \omega_\text{c}$ 
  about the axis of an external magnetic field strength of $1$ T.}
  \item{Fig. 4:
  We simulated (interacting)
  an initially spherical electron SCP
  with $10k$ electrons uniformly distributed in the radius $R_0 = 2 ~\mu$m starting from rest
  and undergoing Coulomb explosion within an external magnetic field strength of $4$ T,
  tuned so that the plasma and cyclotron frequencies are comparable.}
\end{enumerate}

For the envelope equations, the Velocity Verlet algorithm was again employed for integration.
All initial parameters were taken from the initial sample used in the $N$-particle simulations.

\section{Acknowledgements}

Work by R.M.V. and O.Z. was in part supported by a Packard Fellowship in Science and Engineering from the 
David and Lucile Packard Foundation.
Work by P.D. and B.Z. was supported by NSF Grant numbers RC1803719 and RC108666. 

\section{Author contributions}
O.Z. first noticed the phase transition. 
O.Z. and B.Z. developed theoretical model, wrote code, and analyzed the simulation and model data.  
O.Z., R.M.V., P.M.D., and B.Z. interpreted the data and wrote the manuscript.

\section{Data availability}
Data available on request from the authors.  Simulation and envelope code available on request from authors.

\section{Competing interests}
The authors declare no competing interests.

\section{Additional information}
\subsection{Supplementary information} is available for this paper as a separate pdf.

\subsection{Correspondence} Requests for data or code should be addressed to B.Z.
\bibliographystyle{apsrev4-1}
\bibliography{para_dia_osc}

\end{document}